\documentclass[a4paper,11pt]{article}
\usepackage[utf8]{inputenc}
\usepackage{amsfonts}
\usepackage{hyperref}
\usepackage{amssymb}
\usepackage{amstext}
\usepackage{authblk}
\usepackage[pdftex]{graphicx}
\usepackage{cite}

\newcommand{\eq}[1]{\begin{equation} #1 \end{equation}}

\title{Conservation laws for colliding branes with induced gravity}
\author{Mathieu Pellen\thanks{pellen@physik.rwth-aachen.de}}

\affil{Institute for Theoretical Particle Physics and Cosmology, \\
RWTH Aachen University, \\
D-52056 Aachen, Germany}

\begin{document}

\maketitle

\begin{abstract}
We derive conservation laws for collisions of self-gravitating $n$-branes (or $n$-dimensional shells) in an $(n+2)$ dimensional spacetime including induced gravity on the brane.
Previous work has shown how geometrical identities in general relativity enforce conservation of energy-momentum at collisions.
The inclusion of induced gravity terms introduces a gravitational self-energy on the brane which permits energy-momentum conservation of matter fields on the brane to be broken,
so long as the total energy-momentum, including induced gravity terms, is conserved.
We give simple examples with two branes (one ingoing and one outgoing) and three branes.
\end{abstract}

\section{Introduction}

Understanding the hierarchy problem is one the major theoretical challenges in modern physics \cite{Su79,tH80,Ve80}.
It can be formulated in different ways: 
why the corrections to the Higgs boson mass are so big by comparison to its bare mass? 
why there exist very different mass scales in nature? 
why gravity is so weak by comparison to standard model forces?
Indeed the electroweak scale of standard model fields is hundreds of GeV while the Planck scale -where quantum gravity is expected to be dominate- is approximately $10^{19}$ GeV.

To solve this problem, following the original idea of Kaluza \cite{Ka21} and Klein \cite{Kl26}, many models allowing for extra dimensions have been proposed.
Some of these allow large extra dimensions \cite{ArDiDv98,AnArDiDv98} by requiring matter fields to be restricted to lower-dimensional branes while allowing gravity to propagate in a higher-dimensional bulk.
Other studies have attempted to solve the hierarchy problem using warped extra dimensions models \cite{RaSu99_1,RaSu99_2}.
This motivates the study of branes as cosmological objects \cite{BiDeLa99,BiDeElLa99,DvGaPo00}  (for reviews see Ref.~\cite{MaKo10,Wa06,La02}).
In particular, it has been proposed \cite{KhOvStTu01, KaKoLi01} that our universe could be the result of the collision of branes, 
in the so-called ekpyrotic / pyrotechnic scenario.
Thus it is natural to investigate collision of self-gravitating branes (or shells) embedded in higher dimensions 
\cite{DrtH85_1, DrtH85_2, Ne01, LaMaWa02, FrHoSh07, ChKlLe07, BoFr08, EaRiGi09}.
It has previously been shown \cite{LaMaWa02} that for self-gravitating branes in general relativity a simple geometrical identity enforces conservation of energy and momentum at collisions. 

In Ref.~\cite{DvGaPo00}, it was argued that it is possible to generate 4-dimensional Newtonian gravity from a static 3-brane embedded in 5 dimensional Minkowski spacetime by including the effect of induced gravity on the brane,
as might be generated by quantum corrections to classical relativity,
allowing an infinite size flat extra dimension.
The rules derived in Ref.~\cite{LaMaWa02} are general as they rely only on purely geometric considerations \cite{KoIsSe00}.
In particular they account for the fact that the colliding branes modify space-time itself.
It is thus interesting to investigate how these rules are modified when considering induced gravity.

In this paper we will investigate the effect of induced gravity on branes upon energy-momentum conservation rules for branes collisions.
In the first section, we will first recall the derivation of conservation laws for colliding branes in classical general relativity \cite{LaMaWa02}.
After deriving new conservation rules, we will perform a detailed study of the solutions for both the original and modified rules.
In particular, some examples of violation of the principle of matter conservation will be given.

\section{Standard conservation laws}

We consider $n$-dimensional branes living in a $n+2$ dimensional spacetime. 
Let us assume that we have $N$ branes which separate the bulk into $N$ different space-time regions. The dynamics is given by the effective action
\eq{ S = \sum_{i=1}^N \left( S^{(i)}_{EH} + S^{(i)}_{brane} + S^{(i)}_{matter} \right) ,}
where in each bulk space-time region, the Einstein-Hilbert action reads
\eq{ \label{eqn1a}
     S^{(i)}_{EH} = - \frac{1}{2\kappa^2} \int_{B_i}
     {dx^{n+2} \sqrt{-g} \; \cal{R}_{(\mathrm{n+2})}} - 2\Lambda }
and on each brane,
\eq{  \label{eqn1B}
  S^{(i)}_{brane} = - \frac{1}{2\kappa^2} \int_{b_i} {dx^{n+1} \sqrt{-h} \; K}
}
and
\eq{S^{(i)}_{matter} = \int_{b_i}{dx^{n+1} \sqrt{-h} \; \mathcal{L}_{matter}} ,}
where $\cal{R}_{(\mathrm{n+2})}$ is the Ricci scalar in $n+2$ dimensions, $\Lambda$ the cosmological constant, $\kappa^2$ the coupling between matter and gravity,
$\mathcal{L}_{matter}$ the matter Lagrangian, {\it{e.g.}} the standard model, and $K$ the trace of the intrinsic curvature 
associated with the Gibbons-Hawking boundary term \cite{GiHa77} at the brane. 
Note that we do not explicitly include any brane tension in the above and consider it to be part of the matter Lagrangian on the brane.

\begin{figure}[!b]
\begin{center}
\includegraphics[width=5cm]{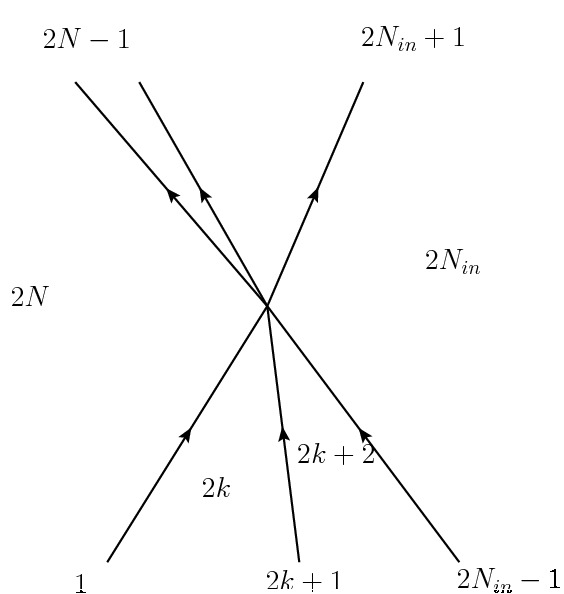}
\end{center}
\caption{Collision of $N_{in}$ ingoing branes giving rise to $N_{out} = N - N_{in}$ outgoing branes.
This is a system of $N$ branes.}
\label{fig5}
\end{figure}

The collision of $N_{in}$ ingoing branes giving rise to $N_{out}$ outgoing branes is shown in Figure~\ref{fig5}.
The bulk regions are empty and can be described by the Schwarzschild-(anti)-de Sitter metric
\begin{equation}
ds^{2}=-f(R)dT^{2}+\frac{dR^{2}}{f(R)}+ R^{2}d\Omega_{n}^{2} ,
\end{equation}
where the ``orthogonal'' metric $d\Omega_{n}^{2} = \gamma_{ij} dx^i dx^j$ is the line-element for a maximally symmetric n-space and does not depend on either $T$ or $R$.
The function $f$ is
\eq{\label{eqn4} f(R) = k - \frac{M}{R^{n-1}} \mp \left(\frac{R}{\ell} \right)^{2} ,}
with $k$ the curvature of the space, $M$ the Schwarzschild mass and $\ell$ the (anti)-de Sitter curvature length.
Thus a brane at the boundary of this region ({\it{i.e.}} the bulk) is described by a two-dimensional trajectory $(T(\tau),R(\tau))$, 
where $\tau$ is the proper time on the brane.
It is then possible to define the two-dimensional velocity vector $u^a = (\dot{T}, \dot{R})$, with dots denoting derivative by respect to the proper time $\tau$.
One can introduce a basis of normalized vectors: ${\bf{e_T}} = f^{-1/2} \frac{\partial}{\partial T}$ and ${\bf{e_R}} = f^{1/2} \frac{\partial}{\partial R}$.
We further define a local Lorentz factor $\gamma = -\bf{e_T}.\bf{u}$ and a relative velocity $\beta$ given by: 
$\gamma \beta = -\bf{e_R}.\bf{u}$.
To characterize the motion of a brane $\mathcal{B}_{2k-1}$  with respect to the static region or bulk $\mathcal{R}_{2k}$, 
we adopted the following definitions for quantities associated with a Lorentz angle:
\eq{\label{eqn5} \gamma_{2k-1|2k} = \cosh(\alpha_{2k-1|2k}) = \sqrt{1+\frac{\dot{R}_{2k-1}^2}{f_{2k}}} }
and
\eq{\label{eqn6} \gamma_{2k-1|2k} \beta_{2k-1|2k} = \sinh(\alpha_{2k-1|2k})  = \epsilon_{2k} \frac{\dot{R}_{2k-1}}{\sqrt{f_{2k}}} ,}
where $\dot{R}$ denotes the derivative of $R$ with respect to $\tau$ and $f$ is the function describing the metric.
Note that $\epsilon$ enables us to fix the convention to draw this situation. If $R$ decreases from ``left'' to ``right'' then $\epsilon = +1$ and $\epsilon = -1$ otherwise. 
Moreover by analogy with special relativity, $\gamma$ and $\beta$ defined above satisfy $\gamma = 1/\sqrt{1-\beta^2}$.
The junction condition \cite{La24,Is66} which represents the jump of extrinsic curvature between two spacetime regions is
\eq{\left[ K_{AB} \right] = - \kappa^2 \left( S_{AB} - \frac{S}{n} g_{AB} \right) ,}
where $S_{AB}$ is the energy-momentum tensor derived from the matter Lagrangian on the brane.
For the orthogonal part, the extrinsic curvature components are $K_{ij} = \left(\epsilon_{2k} / R \right) \sqrt{f_{2k}+\dot{R}^2_{2k-1}} g_{ij}$.
The junction condition becomes
\eq{\epsilon_{2k} \sqrt{f_{2k}+\dot{R}^2_{2k-1}} - \epsilon_{2k-2} \sqrt{f_{2k-2}+\dot{R}^2_{2k-1}} = \frac{\kappa^2}{n} \rho_{2k-1} R,}
with $\rho_{2k-1}$ the comoving energy density associated with the brane $\mathcal{B}_{2k-1}$.
Note the absence of index on $R$; this is due to the fact that $R$ (the radial position)
is the same for any brane meeting at the same point at the same time.
Following Ref.~\cite{LaMaWa02} we define the rescaled brane density as
\eq{\tilde{\rho}_{2k-1} \equiv \pm \frac{\kappa^2}{n} \rho_{2k-1} R,}
with the plus sign for ingoing branes and the minus sign for outgoing branes.

Geometrical consistency~\cite{LaMaWa02} leads to simple conservation rules for the branes. The energy conservation law is
\eq{\sum^N_{k=1} \tilde{\rho}_{2k-1} \gamma_{j|2k-1} = 0,}
while the momentum conservation law is
\eq{\sum^N_{k=1} \tilde{\rho}_{2k-1} \gamma_{j|2k-1} \beta_{j|2k-1} = 0 ,}
for any value of the index j.

\section{Modified conservation laws}

\subsection{Conservation laws}

We now consider the effect of induced gravity on the brane, modifying the standard description.
By induced gravity we mean an Einstein-Hilbert action in $n+1$ dimensions on each $n$-brane.
The assumed action becomes
\eq{ S = \sum_{i=1}^N \left( S^{(i)}_{EH_{(n+2)}} + S^{(i)}_{EH_{(n+1)}} +S^{(i)}_{brane} + S^{(i)}_{matter} \right) ,}
where on each brane
\eq{ \label{eqn1}
     S^{(i)}_{EH_{(n+1)}} = - \frac{1}{2\mu^2} \int_{B_i}
     {dx^{n+1} \sqrt{-h} \; \cal{R}_{(\mathrm{n+1})}},}
with $\cal{R}_{(\mathrm{n+1})}$ the Ricci scalar in $n+1$ dimensions and $\mu^2$ the coupling between matter and gravity.
The junction condition is then:
\eq{[K_{AB}] = - \kappa^2 \left( ( S_{AB} - \frac{S}{n} g_{AB}) + \frac{1}{\mu^2} (\widetilde{U}_{AB} - \frac{\widetilde{U}}{n} g_{AB} ) \right)}
with $\widetilde{U}_{AB}$ the Einstein tensor on the brane ($n+1$ dimensional spacetime) defined as
\eq{\widetilde{U}_{00} = + \frac{n(n-1)}{2} \frac{\delta(\chi)}{\mu^2} (\frac{\dot{R}^2}{R^2} + \frac{k}{R^2})}
and
\eq{\widetilde{U}_{ij} = - (n-1) \frac{\delta(\chi)}{\mu^2} \gamma_{ij} 
\left( R^2 \left( \frac{n-2}{2} \frac{\dot{R}^2}{R^2} + \frac{\ddot{R}}{R} \right) + k \frac{(n-2)}{2} \right).}
Finally by contracting the indices, the junction condition becomes
\eq{ \label{eqn2} \epsilon_{+}\sqrt{f_{+}+\dot{R}^2} - \epsilon_{-}\sqrt{f_{-}+\dot{R}^2} = R \frac{\kappa^2}{n} \rho'_{2k-1} ,}
with the effective energy density defined as
\begin{eqnarray}
\label{eqn7}
\rho'_{2k-1} &=&  \rho_{2k-1}^{matter} - \frac{(n-1)}{2} \frac{n}{\mu^2} (\frac{\dot{R}_{2k-1}^2}{R^2} + \frac{k}{R^2}) \nonumber \\
&=&  \rho_{2k-1}^{matter} - \rho_{2k-1}^{gravity}.
\end{eqnarray}
Here $\rho_{2k-1}^{matter}$ denotes the $\rho_{2k-1}$ of the previous section {\it{i.e.}} the brane energy density.
Recovering the standard case is thus equivalent to set $\rho_{2k-1}^{gravity} = 0$.
Note that this result is in accordance with Ref.~\cite{De01} where the junction condition has been derived for the case $n=3$.
According to this definition it is possible to derive modified conservation laws for energy and momentum for colliding branes
by following the same procedure than in Ref.~\cite{LaMaWa02}.
The modified energy and momentum conservation laws are then:
\eq{\label{eqn8} \displaystyle\sum_{k = 1}^N {\rho'_{2k-1} \gamma_{2k-1|j}} = 0 ,}
and
\eq{\label{eqn9} \displaystyle\sum_{k = 1}^N {\rho'_{2k-1} \gamma_{2k-1|j} \beta_{2k-1|j} } = 0 .}
for any value of the index j\footnote{We here dropped the normalisation factor to simplify notation.
For our purpose, this would not play any role because this normalisation factor would be identical for every branes and thus would always simplify.}.

\subsection{Nil-brane}

We introduce here the concept of nil-brane which will be used abundantly in the following.
By saying nil-brane or vacuum we mean, $\rho' = 0$ and $\rho^{matter} = 0$ which implies $\rho^{gravity} = 0$.
According to the definition of $\rho'$ in equation (\ref{eqn7}), it appears (for $n\ne0,2$) that it gives constraints on the geometry of the bulk.
It implies that for a nil-brane the curvature of the space\footnote{We require that the curvature of the $n$-space must be the same for two regions separated by a brane.} is
\eq{k = - \dot{R}^2 \leq 0 .}
This shows that when nil-branes are considered, the spatial curvature of the $n$-dimensional space will always be negative or zero (in which case $\dot{R}=0$).

\subsection{Diagrammatic description}

\begin{figure}[!b]
\begin{center}
\includegraphics[width=5cm]{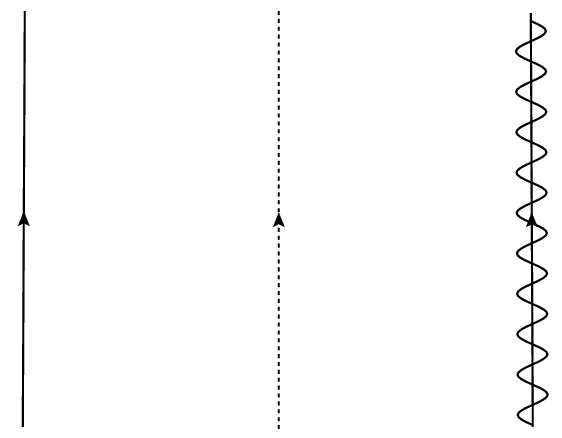}
\end{center}
\caption{Solid lines describe standard branes, dotted lines describe nil-branes and solid-wavy describe branes with $\rho' = 0$ 
but with a non zero matter density.}
\label{fig1}
\end{figure}

In order to simplify the understanding of the different cases that will be discussed in the following, we have adopted a diagrammatic description of the collisions.
As shown in Figure~\ref{fig1}, standard branes will be described by solid lines.
By standard branes, it is understood $\rho' \neq 0$.
Dotted lines will represent nil-branes while solid-wavy lines will represent branes with $\rho' = 0$ 
but with a non zero matter density ({\it{i.e.}} $\rho^{matter} \ne 0$).

\section{Two branes}

We will now focus on particular cases of branes collisions.
The first case studied here is the simplest non trivial one {\it{i.e.}} a two branes system with one ingoing brane and one outgoing brane.
First, the standard case will be studied.
Then the addition of induced gravity will be consider .
It will exhibit new effects, made possible by the use of modified rules.

\subsection{General equations}

The ingoing brane will be denoted by the subscript ``$a$'' while the outgoing brane will be denoted by the subscript ``$b$''.
Form the conservation laws (\ref{eqn8}) and (\ref{eqn9}), one gets\footnote{As the careful reader might have noticed, we have here chosen ``angles'' between two branes and not only between a bulk and a brane.
This is possible by using the fact that $\alpha_{a|b} = \alpha_{a|I} + \alpha_{I|b}$ with $I$ the region between the branes ``$a$'' and ``$b$''.}
\eq{\label{eqn21} \rho'_b = \rho'_a \gamma_{a|b}}
and
\eq{\label{eqn22} \rho'_b \gamma_{b|c} \beta_{b|c} = \rho'_a \gamma_{a|c} \beta_{a|c} .}
To satisfy these equations in general, regardless of the values (non zero) of $\rho'_a$ and $\rho'_b$ the following equation has to be satisfied:
\eq{\label{eqn20} \left( \gamma_{a|b} \gamma_{b|c} \beta_{b|c} - \gamma_{a|c} \beta_{a|c} \right) = 0 .}
It gives a non trivial condition on the expansion rate of the different branes (the expansion rate being defines as: $H = \dot{R}^2 / R^2$).
A particular solution to fulfil this set of equations is to pick $\rho'_b = 0$.
This leads to $\rho'_a = 0$ ($ \gamma_{a|b} \neq 0$ by definition).
We will focus later on this solution.

\subsection{Standard behaviours}

\subsubsection{Standard case}

For the standard case, the definition of the effective energy density is
\eq{\rho' = \rho^{matter},}
{\it{i.e.}} the usual matter density.
The constraint on expansion rate is the same (see equation (\ref{eqn20})).
It has to be stressed that there can be an augmentation of matter in the standard case in the sense that $\rho^{matter}_a$ can be different from $\rho^{matter}_b$.
This difference is ``compensated'' by the motion of the brane.
From equations (\ref{eqn21}) and (\ref{eqn22}), the strict conservation of matter density, $\rho^{matter}_a = \rho^{matter}_b$,
leads to the fact that the trajectories of
the branes ``$a$'' and ``$b$'' are similar.
It means that $\dot{R}_a = \dot{R}_b$ (by definition of $\gamma$, see equation (\ref{eqn5})).
The two branes are indistinguishable.
Thus there is an augmentation of matter density when trajectories evolve and strict conservation if not.

\begin{figure}[!b]
\begin{center}
\includegraphics[width=5cm]{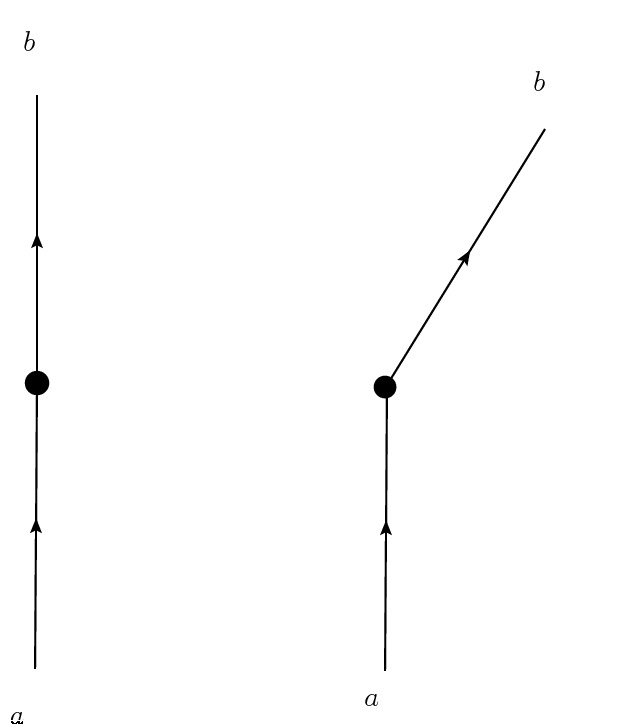}
\end{center}
\caption{On the left hand side, a brane with no change of trajectory and a strict conservation of matter (branes ``$a$'' and ``$b$'' are the same).
On the right hand side, augmentation of matter density through evolution of trajectory.}
\label{fig2}
\end{figure}

What is not possible in the standard case is the creation of matter from nothing {\it{i.e.}} the vacuum.
Indeed, if one considers the case $\rho^{matter}_a =  0$, it necessarily implies $\rho^{matter}_b = 0$.
It means that no matter can be created instantly.
It thus leads to the diagrams showed in Figure~\ref{fig2}.

\subsubsection{Induced gravity case}

In the previous section, it has been shown that the matter density could evolve.
This is still true when induced gravity is considered.
The strict conservation of matter density ($\rho^{matter}_a = \rho^{matter}_b$) leads to a non trivial relation between the ``gravitational'' density and the trajectory namely:
\begin{equation}
 \begin{array}{c}
  \rho^{gravity}_b \gamma_{b|c} \beta_{b|c} (1 - \gamma_{b|c}) -  \rho^{gravity}_a \gamma_{a|c} \beta_{a|c} (1 - \gamma_{a|b}) = \\
  \rho^{gravity}_a ( \gamma_{b|c} \beta_{b|c} - \gamma_{a|c} \beta_{a|c} ) - \rho^{gravity}_b \gamma_{a|b} ( \gamma_{b|c} \beta_{b|c} - \gamma_{a|c} \beta_{a|c} ).
 \end{array}
\end{equation}

If one now considers the case where $\rho'_a = 0$ implying $\rho'_b = 0$, the conservation laws become
\eq{\label{eqn14} \rho_{a}^{matter} - \frac{(n-1)}{2} \frac{n}{\mu^2} (\frac{\dot{R}_{a}^2}{R^2} + \frac{k}{R^2}) =
\rho_{b}^{matter} - \frac{(n-1)}{2} \frac{n}{\mu^2} (\frac{\dot{R}_{b}^2}{R^2} + \frac{k}{R^2}) = 0.}
Demanding a strict preservation of matter density leads to:
\eq{\label{eqn15} \rho^{gravity} = \frac{(n-1)}{2} \frac{n}{\mu^2} (\frac{\dot{R}_{a}^2}{R^2} + \frac{k}{R^2})
= \frac{(n-1)}{2} \frac{n}{\mu^2} (\frac{\dot{R}_{b}^2}{R^2} + \frac{k}{R^2}) .}
This implies (if $\mu$ is assumed to be constant) that $\dot{R}_{a} = \dot{R}_{b}$ meaning that both branes have the same velocity {\it{i.e.}} they are identical .

For now we retain that the introduction of induced gravity still allows strict conservation of matter density as well as its augmentation.
The conditions for the evolution of matter density or its strict conservation are more complicated in this case.
Nonetheless, it still gives the possibility to recover regular behaviours from the standard case (these cases are described by Figure~\ref{fig2}).

\subsection{A new mechanism for the evolution of matter density}

The previous section described cases of matter evolution with and without induced gravity.
We will now study the cases where the introduction of induced gravity gives the opportunity to create matter instantly.
Indeed according to the new energy/momentum conservation laws, we identified a new mechanism that allows for the instant creation of a single brane from a nil-brane.
As pointed out previously, one solution is to pick $\rho'_a = \rho'_b = 0$.
Thus in this section, the effective density $\rho'$ will always be assumed to be null.
This mechanism relies on the possibility to have a change in the trajectory (possibly infinitesimal).
For the sake of clarity we recall that the setting studied is encompassed in the following equations (see equation (\ref{eqn14})):
\eq{\label{eqn23} \rho_{a}^{matter} - \frac{(n-1)}{2} \frac{n}{\mu^2} (\frac{\dot{R}_{a}^2}{R^2} + \frac{k}{R^2}) =
\rho_{b}^{matter} - \frac{(n-1)}{2} \frac{n}{\mu^2} (\frac{\dot{R}_{b}^2}{R^2} + \frac{k}{R^2}) = 0.}
In the case of a null effective energy density the junction condition (see equation (\ref{eqn2})) becomes
\eq{ \label{eqn3} \epsilon_{+}\sqrt{f_{+}+\dot{R}^2} - \epsilon_{-}\sqrt{f_{-}+\dot{R}^2} = 0 .}
According to this equation two sub-cases appear: the so-called symmetric case ($\epsilon_+.\epsilon_+=1$) 
and the orbifold case ($\epsilon_+.\epsilon_+=-1$ and $f_+ = f_-$)\footnote{Note that equation (\ref{eqn3}) necessarily lead to identical spacetime on both side of the brane.
Indeed for nil-branes, it is understandable that ``both sides'' are identical because a nil-brane is the vacuum.}.
We will see that both cases do not necessarily lead to instant creation of matter.
An important point in this section is that the incoming brane ``$a$'' will always be a nil-brane, meaning $\rho'_{a} = \rho_{a}^{matter} = 0$.
The conservation equations becomes (see equation (\ref{eqn23})):
\eq{\label{eqn25} \frac{(n-1)}{2} \frac{n}{\mu^2} (\frac{\dot{R}_{a}^2}{R^2} + \frac{k}{R^2}) =
\rho_{b}^{matter} - \frac{(n-1)}{2} \frac{n}{\mu^2} (\frac{\dot{R}_{b}^2}{R^2} + \frac{k}{R^2}) = 0.}
No change of gravity is assumed here that is to say, the gravitational constant $\mu$ is assumed to be constant at the collision point.

\subsubsection{Symmetric case}

\begin{figure}[!b]
\begin{center}
\includegraphics[width=2.5cm]{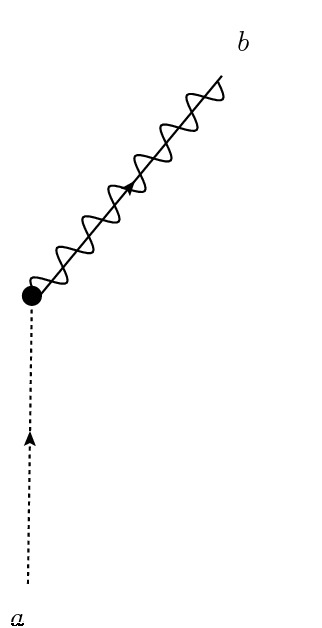}
\end{center}
\caption{A nil-brane ``$a$'' giving rise to a brane ``$b$'' where $\rho'_b = 0$ and $\rho^{matter}_b \neq 0$ {\it{i.e.}} a case of instant creation of matter.}
\label{fig3}
\end{figure}

By symmetric case we mean the setting where $\epsilon_{-} \epsilon_{+} = +1$.
By looking at equation (\ref{eqn3}), the equation describing the ``boundary condition'' of the two regions surrounding a given brane, one obtains
\eq{f_{+}(R_{brane}) = f_{-}(R_{brane}) \; \forall \; R_{brane} ,}
with $R_{brane}$ the trajectory of the brane.
This being true for any point of the trajectory then it gives: $f'_{+}(R_{brane}) = f'_{-}(R_{brane})$ with 
$f'$ the derivative of the function $f : R \rightarrow f(R)$.
Thus if one performs a Taylor expansion of $f_{+/-}(R)$, for $R$ sufficiently close to the brane trajectory $R_{brane}$, it gives:
\eq{f_{+/-}(R) = f_{+/-}(R_{brane}) + f'_{+/-}(R_{brane}) \left( R-R_{brane} \right) .}
Thus, it can be inferred that:
\eq{f_{+}(R) = f_{-}(R) \; \forall \; R .}
It means that the spacetime is identical on both sides of the brane.
Moreover the conservation rules (see equation (\ref{eqn25})) gives
\eq{ \rho^{matter}_b = \frac{n(n-1)}{2 \mu^2 R^2} \left( \dot{R}_{b}^2 - \dot{R}_{a}^2 \right) \neq 0.}
Hence it appears that the creation of matter on the outgoing brane is due to the change of velocity of the brane.
Recall that for the incoming brane, it was assumed $\rho^{matter}_a = 0$ {\it{i.e.}} there was no matter on it.
Thanks to this new conservation laws, it is now possible to generate a non zero matter density on the outgoing brane through a change of trajectory 
at the ``collision point''.
The outgoing brane is no more a nil-brane (see Figure~\ref{fig3}).

\subsubsection{Orbifold case}

\begin{figure}[!b]
\begin{center}
\includegraphics[width=3cm]{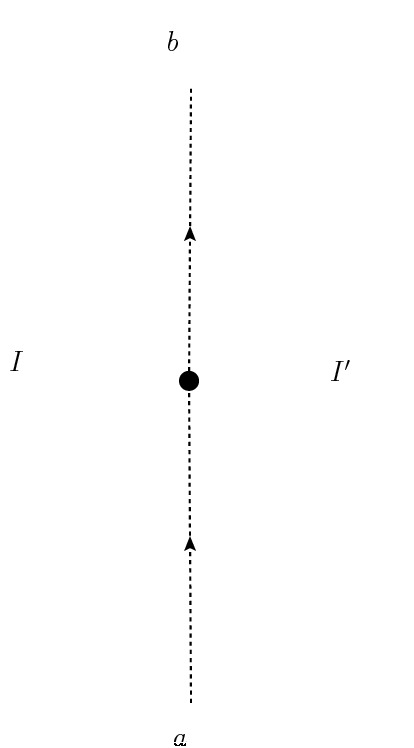}
\end{center}
\caption{A $\mathbb{Z}_2$ symmetric brane will always respect a strict conservation of matter density and its trajectory will never evolve.}
\label{fig4}
\end{figure}

Note that this setting will not lead to creation of matter when one considers only two branes.
The orbifold case stands in this context for $\epsilon_{-} \epsilon_{+} = -1$ and $f_{+} = f_{-}$ (spacetime are identical on both side).
Here, equation (\ref{eqn3}) gives
\eq{f(R)+\dot{R}^2 = 0 \; .}
According to equations (\ref{eqn4}) and (\ref{eqn1}), it gives a ``new'' definition of $\rho^{gravity}$:
\eq{\rho^{gravity} = \frac{(n-1)}{2} \frac{n}{\mu^2} \frac{1}{R^2} \left( \frac{M}{R^{n-1}} \pm \left(\frac{R}{l}\right)^2 \right)  .}
Knowing that $\rho^{gravity}_a = 0$ (nil-brane for incoming brane), it necessarily leads to $\rho^{gravity}_b = 0$ 
because  $\rho^{gravity}$ will not change at the collision point.
Indeed $\rho^{gravity}$ depends only on $R$.
$R$ being the position of the collision, there are no possibilities for a modification of the ``gravitational'' density.
Note that the variation of $\mu$ would still lead to  $\rho^{matter}_b = 0$.
Thus in the case of $\mathbb{Z}_2$ symmetric two branes system, it is impossible to create instantly 
single brane (and thus matter) from the vacuum ({\it{i.e.}} nil-brane).
The $\mathbb{Z}_2$ symmetry prevent instant creation of matter from vacuum.
More than that, it ensures a strict conservation of matter density and the trajectory will not evolve (see Figure~\ref{fig4}).

\section{Three branes}

\begin{figure}[!b]
\begin{center}
\includegraphics[width=4cm]{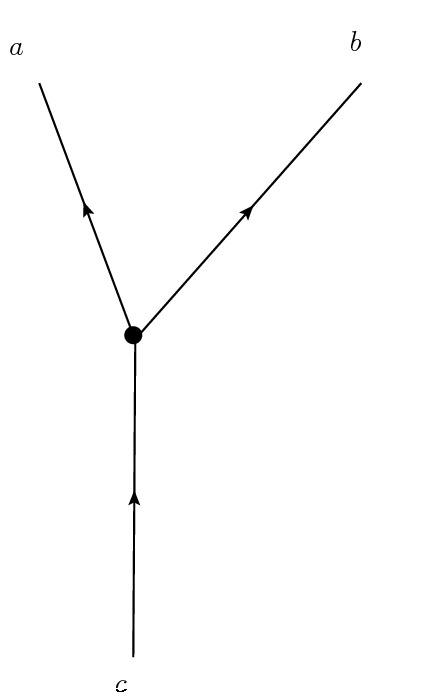}
\end{center}
\caption{One ingoing brane yielding two outgoing branes in the standard case.}
\label{fig6}
\end{figure}

This section is devoted to the study of three branes system.
Here the ingoing brane will be denoted by the subscript ``$c$'' while the two outgoing branes will be respectively denoted by the subscript ``$a$'' and ``$b$''.
This situation is described in Figure~\ref{fig6}.
We will follow the same procedure as before by first reviewing the case 
where there is no violation of matter conservation and no brane creation in both standard and modified case.

\subsection{General equations}

In a frame in which the ``$c$'' brane is stationary, the momentum and energy conservation laws (see equations (\ref{eqn8}) and (\ref{eqn9})) give:
\eq{\label{eqn10} 0 = \rho'_a \gamma_{a|c} \beta_{a|c} + \rho'_b \gamma_{b|c} \beta_{b|c} ,}
and
\eq{\label{eqn11} \rho'_c = \rho'_a \gamma_{a|c} + \rho'_b \gamma_{b|c}.}
These equations thus give a non trivial condition between the effective energy density and the expansion rate of the three branes.

By considering the case where the effective energy density of the incoming brane is null {\it{i.e.}} $\rho'_c = 0$, one obtains
\eq{0 = \rho'_a \gamma_{a|c} \beta_{a|c} + \rho'_b \gamma_{b|c} \beta_{b|c} ,}
and
\eq{\label{eqn24} \rho'_a \gamma_{a|c} + \rho'_b \gamma_{b|c} =0 .}
Given that $\gamma\neq0$, it implies
\eq{\rho'_b \left( \beta_{a|c} - \beta_{b|c} \right) = 0 .}
Finally, either brane ``$a$'' and ``$b$'' have the same trajectory 
({\it{i.e.}}, $\beta_{a|c}=\beta_{b|c}$ and hence $\dot{R}_a=\dot{R}_b$) and $\rho'_a = - \rho'_b$ or $\rho'_a = \rho'_b =0$.
Having the same trajectory means that they merge.
Thus this merged brane have an effective energy density $\rho'_{ab} = \rho'_{a} + \rho'_{b} = 0$.
It corresponds to the system of a 2 branes previously studied.
Thus for a three branes system, if the incoming brane has a null effective energy density, then the outgoing branes have both a null effective energy density
or merge into a brane of null effective energy density.

\begin{figure}[!b]
\begin{center}
\includegraphics[width=3cm]{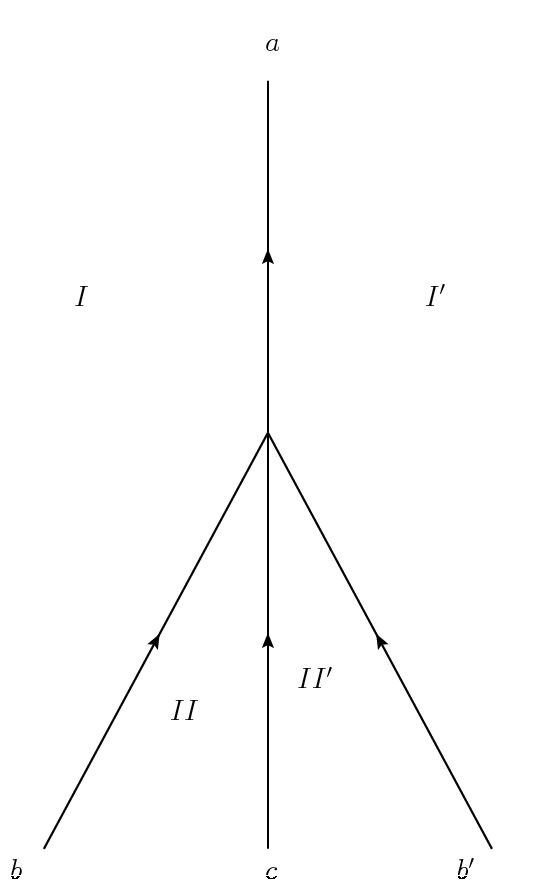}
\end{center}
\caption{A $\mathbb{Z}_2$ symmetric brane ``$c$'' collides a ``normal'' brane ``$b$'' to give rise to a $\mathbb{Z}_2$ symmetric brane ``$a$''.}
\label{fig8}
\end{figure}

The setting for a three branes system in the orbifold case is: one incoming $\mathbb{Z}_2$ brane ``$c$'' and one normal incoming brane ``$b$'' are considered.
The outgoing brane ``$a$'' is also considered $\mathbb{Z}_2$ symmetric\footnote{Just as a remind, due to the $\mathbb{Z}_2$ symmetry of branes ``$a$'' and ``$c$'', $\alpha_{a|c}=0$.}.
The equations are:
\eq{\label{eqn18} 0 = \rho'_b \gamma_{b|c} \beta_{b|c} }
and
\eq{\label{eqn19} \rho'_c = \rho'_a + 2 \rho'_b \gamma_{b|c}.}
This case is displayed in Figure~\ref{fig8}.

\subsection{Standard behaviours}

In this part, it will be shown that modified conservation laws can also mimic the behaviours of three branes systems with standard conservation laws.

\subsubsection{Standard case}

As explained previously, in the standard case the contribution from the induced gravity is null.
Thus equations (\ref{eqn10}) and (\ref{eqn11}) becomes:
\eq{ 0 = \rho^{matter}_a \gamma_{a|c} \beta_{a|c} + \rho^{matter}_b \gamma_{b|c} \beta_{b|c} ,}
and
\eq{ \rho^{matter}_c = \rho^{matter}_a \gamma_{a|c} + \rho^{matter}_b \gamma_{b|c}.}
Of course, as for the two branes systems, there can be augmentation of matter {\it{i.e.}} $\rho_c < \rho_a + \rho_b$.
This augmentation being compensated by the dynamics of the branes.
If one now demands a strict conservation of matter density {\it{i.e.}} $\rho_c = \rho_a + \rho_b$, it leads to
\eq{\gamma_{a|c} \beta_{a|c} (1-\gamma_{a|c}) = \gamma_{b|c} \beta_{b|c} (1-\gamma_{b|c}) .}
This means that the trajectories of ``$a$'' and ``$b$'' are similar {\it{i.e.}} there is just one outgoing brane.
This is thus the two branes case and the conclusion still hold:
if there is strict conservation of matter density, there is no change of trajectory at the collision point.

Finally if one considers that the ingoing branes has $\rho^{matter}_c = 0$ -thanks to our previous calculation- it implies $\rho^{matter}_a = \rho^{matter}_b = 0$.
Thus -as usual in the standard case- instant creation of matter is impossible.
The conclusion for three branes systems are thus similar to the one for two branes systems.

\subsubsection{Induced gravity case}

By considering induced gravity, equations (\ref{eqn10}) and (\ref{eqn11}) thus becomes

\eq{\label{eqn12}
    \begin{array}{l}
    0 = \left( \rho_{a}^{matter} - \frac{(n-1)}{2} \frac{n}{\mu^2} (\frac{\dot{R}_{a}^2}{R^2} + \frac{k}{R^2}) \right) \gamma_{a|c} \beta_{a|c} \\
    + \left( \rho_{b}^{matter} - \frac{(n-1)}{2} \frac{n}{\mu^2} (\frac{\dot{R}_{b}^2}{R^2} + \frac{k}{R^2}) \right) \gamma_{b|c} \beta_{b|c} ,
    \end{array}
}
and
\eq{\label{eqn13}
    \begin{array}{l}
    \left( \rho_{c}^{matter} - \frac{(n-1)}{2} \frac{n}{\mu^2} (\frac{\dot{R}_{c}^2}{R^2} + \frac{k}{R^2}) \right) \\
    = \left( \rho_{a}^{matter} - \frac{(n-1)}{2} \frac{n}{\mu^2} (\frac{\dot{R}_{a}^2}{R^2} + \frac{k}{R^2}) \right) \gamma_{a|c}  \\
    + \left( \rho_{b}^{matter} - \frac{(n-1)}{2} \frac{n}{\mu^2} (\frac{\dot{R}_{b}^2}{R^2} + \frac{k}{R^2}) \right) \gamma_{b|c}.
    \end{array}
}
We will not perform an extensive study of this case.
Indeed, no new effects or mechanisms concerning the strict conservation of matter density ($\rho_c = \rho_a + \rho_b$) have been observed.
This statement holds when $\rho'_c \neq 0$ and $\rho'_c = 0$.
Again, if an ingoing branes has $\rho'_c = 0$, it implies $\rho'_a = \rho'_b = 0$.

\subsubsection{New behaviours}

We will proceed as for the two branes system cases.
As previously, the incoming brane ``$c$'' will be a nil-brane and ``$a$'' and ``$b$'' the outgoing branes.
It will again be assumed that $\rho'_c = \rho^{matter}_c = 0 = \rho^{gravity}_c$.

\subsubsection{Symmetric case}

\begin{figure}[h!]
\begin{center}
\includegraphics[width=4cm]{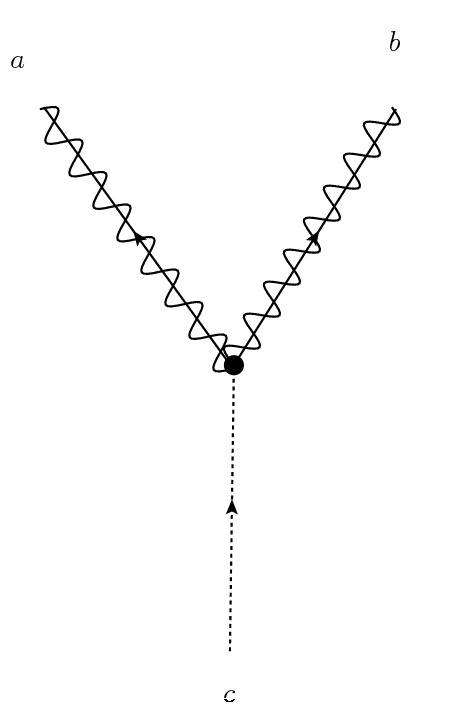}
\caption{A nil-brane ``$c$'' giving rise to two  zero effective density branes ``$a$'' and ``$b$''.}
\label{fig7}
\end{center}
\end{figure}

As previously explained, assuming $\rho'_c =0$ implies that $\rho'_a = \rho'_b = 0$.
By looking at equations (\ref{eqn12}) and (\ref{eqn13}), one obtains:
\eq{\label{eqn16} \rho'_a = \rho^{matter}_a -  \frac{(n-1)}{2} \frac{n}{\mu^2} \frac{1}{R^2} \left( \dot{R}_{a}^2 - \dot{R}_{c}^2 \right) = 0 ,}
and
\eq{\label{eqn17} \rho'_b = \rho^{matter}_b -  \frac{(n-1)}{2} \frac{n}{\mu^2} \frac{1}{R^2} \left( \dot{R}_{b}^2 - \dot{R}_{c}^2 \right) = 0.}
It is clear that if the trajectory of the ingoing brane is different from the trajectory of one of the outgoing branes,
the matter density of one of the outgoing brane will be non zero.
This situation is depicted on Figure~\ref{fig7}.
It means that if the outgoing branes do not have the same trajectory, there will necessary be instant creation of matter.
This constitutes a stronger statement than before.
If nil-branes exists and can give rise to two distinctive branes then there will necessarily be instant creation of matter.
Note again that if the trajectories of the two outgoing branes are similar, the situation is described by the two branes system.
Same trajectory being here interpreted as a merging of the two branes.

\subsection{Orbifold case}

\begin{figure}[h!]
\begin{center}
\includegraphics[width=4cm]{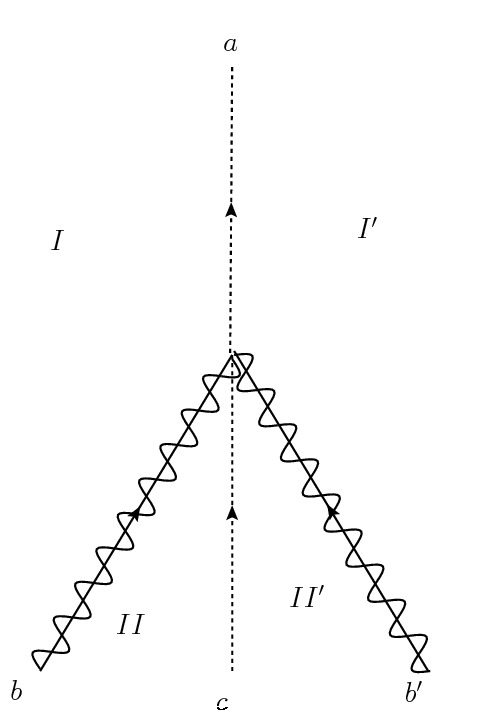}
\caption{A $\mathbb{Z}_2$ symmetric nil-brane ($c$) collides with a brane ($b$) with null effective energy density but with non zero matter density to give rise to a $\mathbb{Z}_2$ symmetric nil-brane ($a$).}
\label{fig10}
\end{center}
\end{figure}

In this setting, the incoming $\mathbb{Z}_2$ brane is denoted by ``$c$'' and the normal incoming brane is denoted by ``$b$''.
The outgoing brane ``$a$'' is also considered as $\mathbb{Z}_2$ symmetric (see Figure~\ref{fig10}).
The equations for the conservation laws are (\ref{eqn18}) and (\ref{eqn19}).
They give $\rho'_c = \rho'_a$ {\it{i.e.}} the effective matter density for the two $\mathbb{Z}_2$ branes are identical.
It has been shown previously (see equation (\ref{eqn4})) that the ``gravitational'' density $\rho^{gravity}$ 
of a $\mathbb{Z}_2$ brane is entirely determined by the parameters at the collision point.
In particular, when collisions of only $\mathbb{Z}_2$ branes are considered,
$\rho^{gravity}$ has to be identical for every $\mathbb{Z}_2$ brane at the collision point.
Thus when collisions including $\mathbb{Z}_2$ branes are considered, all the $\mathbb{Z}_2$ branes have necessarily the same matter density.
In the case studied here with one incoming and one outgoing $\mathbb{Z}_2$ brane, 
there will necessarily be conservation of matter between those two branes\footnote{We did not study a system of more that three $\mathbb{Z}_2$ branes because one dimension of spacetime would disappear at the collision.}.

Equations (\ref{eqn18}) and (\ref{eqn19}) imply: $\rho'_b = \rho^{matter}_b - \rho^{gravity}_b = 0$.
Nothing constraints the value of the ``gravitational'' density of the normal brane.
In particular, $\rho^{matter}_b$ can take any value.
This constitutes an instant decay of matter disappearing into the vacuum\footnote{Note that ``time'' can be reversed and the normal incoming brane can become an outgoing brane.
The instant decay would become an instant creation of matter.} (see Figure~\ref{fig10}).
Note that if another ``normal'' brane is added, the conclusion would still hold {\it{i.e.}} no instant creation of matter 
between the two $\mathbb{Z}_2$ (a strict conservation) but possibly for the two (or more) non-$\mathbb{Z}_2$ branes.

\section{Conclusion}

We have presented here a detailed study of conservation laws for colliding $n$-branes (or $n$-dimensional shells) in arbitrary dimension.
After reviewing the standard rules, we have derived new conservation rules when considering self-gravitating $n$-branes by including induced gravity on the brane.
We have exhibited various simple examples with two or three branes for both the standard case and the induced gravity case.
These examples showed that the inclusion of induced gravity leads to drastically different behaviours.
In particular these examples shed light on a possible breakdown of energy-momentum conservation violating the principle of matter conservation.
We have also showed that the $\mathbb{Z}_2$ symmetry plays a special role in this context.
Indeed, when considering systems involving $\mathbb{Z}_2$ symmetric branes, it automatically ensures the strict conservation of matter density.

We have found a mechanism where a change in the trajectory of a brane could induce the instant creation of single branes from the vacuum.
Thus, the existence of nil-branes in this context leads -in general- to a violation of matter conservation.
This could in particular lead to unconstrained instant creation of $n$-branes from the vacuum.
Despite the fact we have found a symmetry preventing this, having solutions violating the principle of matter conservation shows that these rules are not sufficient.
Excitation of the vacuum is a common behaviour in quantum theory but we have made here a classical calculation.
Knowing that induced gravity terms originate from quantum corrections, this would indicate that quantum conservation rules are required.

\section*{Acknowledgements}

I would like to thank D. Wands for his supervision and comments, M. Kr\"{a}mer for suggestions, the ICG of Portsmouth for its hospitality and J. Emery \& T. Kidani for fruitful discussions.
Diagrams have been drawn using JaxoDraw \cite{BiTh03}.

% Working
% \bibliography{Article}
% \bibliographystyle{ieeetr}

% Not working
% \bibliographystyle{plain}
% \input{Article.bbl}

% Not working
  \bibliographystyle{spmpsci}
  \bibliography{Article}

\end{document}